# Parasitic erbium photoluminescence in commercial telecom fiber optical components

Gary Wolfowicz[1,*], F. Joseph Heremans[1,2], David D. Awschalom[1,2,3]

[1]*Center for Molecular Engineering and Materials Science Division, Argonne National Laboratory, Lemont, IL 60439, USA*

[2]*Pritzker School of Molecular Engineering, University of Chicago, Chicago, IL 60637, USA*

[3]*Department of Physics, University of Chicago, Chicago, IL 60637, USA*

*gwolfowicz@photonic.com

**Noiseless optical components are critical for applications ranging from metrology to quantum communication. Here we characterize several commercial telecom C-band fiber components for parasitic noise using a tunable laser. We observe the spectral signature of trace concentrations of erbium in all devices from the underlying optical crystals including $YVO_4$, $LiNbO_3$, $TeO_2$ and AMTIR glass. Due to the long erbium lifetime, these signals are challenging to mitigate at the single photon level in the telecom range, and suggests the need for higher purity optical crystals.**

Today, most optical fiber components are designed around the optical telecom S-, C- and L-bands where the minimum absorption loss in optical fibers occurs. In particular, the C-band between 1527 nm and 1565 nm is most commonly used as it is matched with the emission of erbium ions, which provides an efficient gain medium for lasers and optical amplifiers. More recently, erbium ions have also been investigated for quantum communication applications [1–3]. Erbium ions are particularly attractive as they have a very long excited lifetime (1-10 ms typical) for efficient population inversion and can be easily incorporated in most transparent glasses with fairly constant absorption and emission wavelength [4,5].

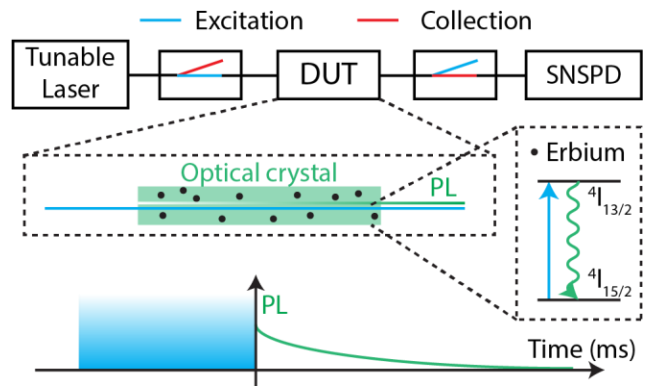

**Fig. 1. Device characterization setup for telecom optical components.** A device-under-test (DUT) is sequentially excited by a tunable laser and then measured using a superconducting nanowire single photon detector (SNSPD). The device emits photoluminescence (PL) with slow decay from the parasitic erbium impurities within the internal optical crystal.

In parallel, very low light applications reliant on single-photon detection such as photon correlation spectroscopy and quantum optics, as well as single molecule or single defect microscopy, demand extremely weak background noise levels typically below kcounts/s or equivalent fW optical power [6,7]. Since trace amounts of rare-earth impurities in the part-per-million range is common in many optical grade glasses within active fiber components [8], these unintended emitters often become a relevant source of noise. Furthermore, the same desired optical properties of long lifetime and telecom wavelength make the erbium ions extremely difficult to distinguish spectrally or temporally from an actual measured signal.

In this letter, we investigate the parasitic erbium emissions in several typical fiber components, including switches, a circulator, and modulators. We find that erbium photoluminescence is noticeably present at the fW level in most of these devices, up to a few orders of magnitudes higher than the dark count (10-100 counts/s or 1-10 aW at 1530 nm) of a superconducting nanowire single photon detector (SNSPD). The combination of spectra and lifetime can provide a clear signature to identify the underlying optical material, and potentially quantify the impurity concentration.

Our all-fiber C-band characterization apparatus, shown in Fig. 1, consists of a narrow-line tunable laser (Pure Photonics), a series of polarization-maintaining acousto-optic (A-O) modulators (AA Opto-Electronic) for ultra-high extinction ratio (>150 dB) switching, the fiber device-under-test (DUT) and a single A-O modulator protecting a high quantum efficiency (~80%) broadband telecom (1200-1600nm) SNSPD (Quantum Opus). The laser power is set to 1 mW at the DUT input. All signals are corrected from any background signal without the DUT in the setup.

We consider two types of experiments: resonant excitation spectra and lifetime measurements. In both cases, the laser is pulsed for 3 ms using the A-O modulators and excites the DUT, while the SNSPD is protected by an A-O modulator during this phase. The excitation is then turned off and the photoluminescence is detected by the SNSPD after a short delay (0.1 ms). For the resonant spectra, the tunable laser is swept across its C-band range and the signal integrated. For the lifetime experiments, the laser is set to a fixed wavelength (at the highest intensity peak) while the transient decay is measured.

We first test three-port optical components based on yttrium vanadate ($YVO_4$) birefringent crystals including a circulator (CIR-PM-15, AFW Technologies) and a fast single pole double throw (1x2) switch (CLBD-125311313, Agiltron). The results are compared to a reference unintentionally doped optical-grade $YVO_4$ crystal (MTI Corporation) in a separate free-space confocal microscope. These three-port optical devices are commonly used for routing light (e.g. excitation, collection) or for protecting sensitive components (e.g. lasers, detectors).

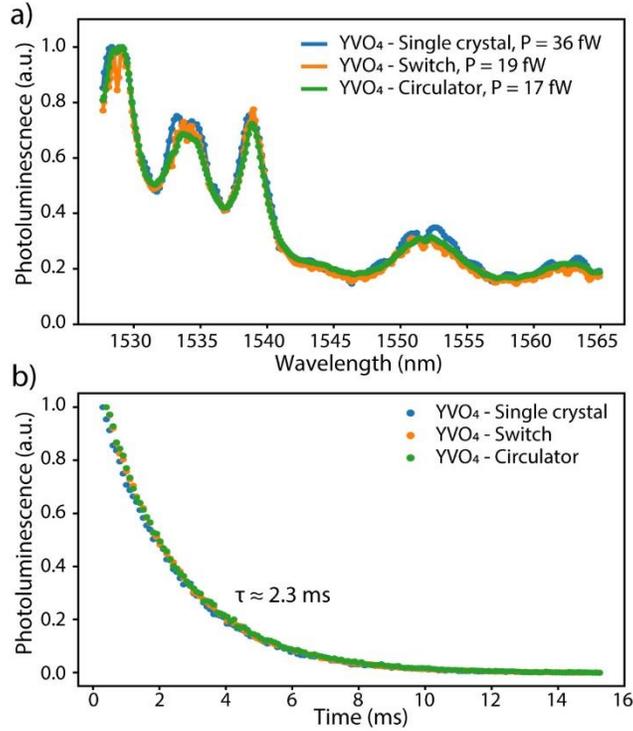

**Fig. 2. YVO$_4$-based optical components.** a) C-band spectra for an optical switch, a circulator and a bulk single crystal as reference. In the legend, P is the measured equivalent power for the highest intensity peak (at 1529 nm). b) Corresponding excited state decay times at 1529 nm.

In Fig. 2.a), we show the photoluminescence under resonant excitation as the tunable laser wavelength is swept, and in b) the corresponding excited state lifetime measured at the highest peak (1529 nm). The spectrum and lifetime combination, which overlap for both components with the reference YVO$_4$ crystal, provide a very strong signature matching that of Er:YVO$_4$ in the literature [9,10]. This includes three main peaks at 1529 nm, 1534 nm and 1538.8 nm and a lifetime about 2.3 ms. The highest peak photoluminescence is observed for the switch with ~150 kcounts/s, equivalent to ~19 fW power. By comparison, the erbium emission from the reference undoped YVO$_4$ is also only ~280 kcounts/s, though with different optical depth in the free-space setup, yet shows that such optical components could contribute significantly as a noisy background for material characterization. Finally, despite these devices having three ports normally separated with large cross-talk attenuation (>50 dB), the photoluminescence is found to be present and equal in all combination of input and output ports as it is emitted internally. The resulting erbium noise could therefore leak into a sensitive optical path.

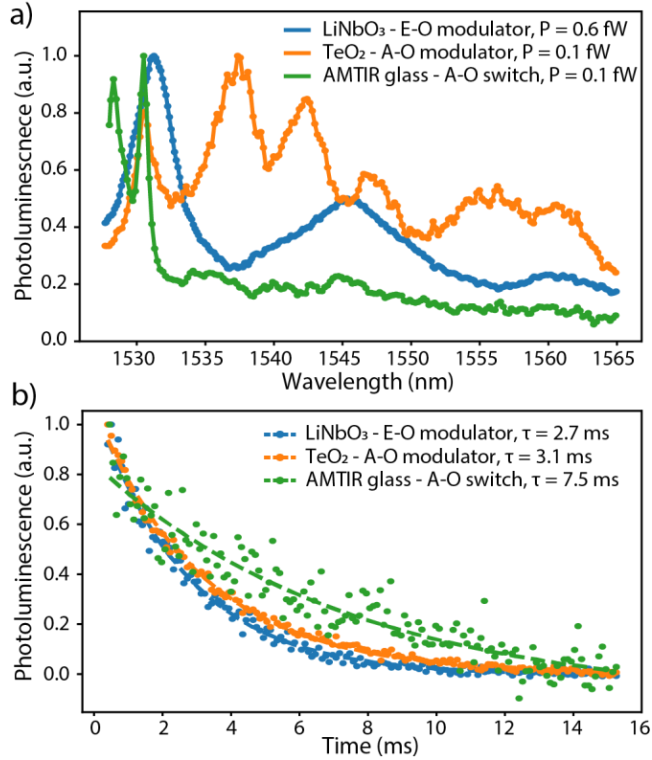

**Fig. 3. Fast optical modulators.** a) C-band spectra for a LiNbO$_3$ electro-optic (E-O) modulator, a TeO$_2$ acousto-optic (A-O) modulator and a AMTIR glass A-O nanosecond switch. In the legend, P is the measured equivalent power for the highest intensity peak. b) Corresponding excited state decay times with an exponential decay fit in dashed lines.

The second group of optical components characterized here are fast (MHz-GHz) optical modulators based on electro-optic (E-O) and A-O crystals. This includes an E-O modulator based on LiNbO$_3$ (MPZ-LN-10-00, iXblue), an A-O modulator with a TeO$_2$ crystal (MT80-IIR30-Fio-PM0, AA Opto-Electronic), and an A-O single pole double throw (1x2) switch based on a chalcogenide AMTIR glass (OS-2-1-C-100, Brimrose). In Fig. 3), we characterize again both the wavelength excitation response and the transient decay from the excited state, providing unique identifiers for noise source in these components. The spectra indeed match erbium in LiNbO$_3$ [11] and coarsely (broadened spectrum) TeO$_2$ [12] in literature, however we could not obtain a reference for the AMTIR glass whose long lifetime and spectrum still suggest erbium emission. It should be noted that the equivalent power for these devices is very low (sub-fW), slightly above the limit of our SNSPD detector, and would therefore be only relevant in the most stringent cases. On the other hand, the sharp peaks from the AMTIR-based switch could easily be mistaken, for example, for the spectral signature of a sample in a characterization setup.

In conclusion, we find that trace impurities of erbium ions can result in non-negligeable amount of background resonant noise for applications at the single photon level, especially within three-ports components such as circulators which are often used to protect a detection path. In particular, YVO$_4$-based optical components have been measured to have the highest photon count across multiple components and suppliers. This work therefore suggests the need for higher optical grade quality and awareness of the presence of erbium impurities in optical components.

**Funding.** This work and all authors were primarily supported by the U.S. Department of Energy, Office of Science, Basic Energy Sciences, Materials Sciences and Engineering Division.

**Acknowledgments.** We thank S. Sullivan, A. Dibos, T. Zhong and C. Wicker for fruitful discussions.

**Disclosures.** The authors declare no conflicts of interest.

**Data Availability.** The data that support the findings of this Letter are available from the corresponding author upon reasonable request.

## References

1. T. Zhong and P. Goldner, "Emerging rare-earth doped material platforms for quantum nanophotonics," Nanophotonics **8**, 2003–2015 (2019).

2. J. M. Kindem, A. Ruskuc, J. G. Bartholomew, J. Rochman, Y. Q. Huan, and A. Faraon, "Control and single-shot readout of an ion embedded in a nanophotonic cavity," Nature **580**, 201–204 (2020).

3. A. M. Dibos, M. Raha, C. M. Phenicie, and J. D. Thompson, "Atomic Source of Single Photons in the Telecom Band," Phys. Rev. Lett. **120**, 243601 (2018).

4. P. Nekvindova, A. Mackova, and J. Cajzl, "Erbium luminescence in various photonic crystalline and glass materials - A review," Int. Conf. Transparent Opt. Networks 2–6 (2017).

5. W. J. Miniscalco, "Erbium-Doped Glasses for Fiber Amplifiers at 1500 nm," J. Light. Technol. **9**, 234–250 (1991).

6. G. Wolfowicz, C. P. Anderson, B. Diler, O. G. Poluektov, F. J. Heremans, and D. D. Awschalom, "Vanadium spin qubits as telecom quantum emitters in silicon carbide," Sci. Adv. **6**, eaaz1192 (2020).

7. A. T. K. Kurkjian, D. B. Higginbottom, C. Chartrand, E. R. MacQuarrie, J. R. Klein, N. R. Lee-Hone, J. Stacho, C. Bowness, L. Bergeron, A. DeAbreu, N. A. Brunelle, S. R. Harrigan, J. Kanaganayagam, M. Kazemi, D. W. Marsden, T. S. Richards, L. A. Stott, S. Roorda, K. J. Morse, M. L. W. Thewalt, and S. Simmons, "Optical observation of single spins in silicon," arXiv:2103.07580 (2021).

8. A. P. D'Silva and V. A. Fassel, "X-Ray Excited Optical Fluorescence of Trace Rare Earths in Yttrium Phosphate and Yttrium Vanadate Hosts The Part per Giga Level Determination of Rare Earth Impurities in Yttrium Oxide," Anal. Chem. **45**, 542–547 (1973).

9. N. Ter-Gabrielyan, V. Fromzel, W. Ryba-Romanowski, T. Lukasiewicz, and M. Dubinskii, "Spectroscopic and laser properties of resonantly (in-band) pumped Er:YVO4 and Er:GdVO4 crystals: a comparative study," Opt. Mater. Express **2**, 1040 (2012).

10. S. Golab, G. Dominiak-Dzik, P. Solarz, T. Lukasiewicz, M. Swirkowicz, I. Sokolska, and W. Ryba-Romanowski, "Relaxation dynamics of excited states of Er3+ in YVO4 single crystals," Growth, Charact. Appl. Single Cryst. **4412**, 380–384 (2001).

11. M. Mattarelli, S. Sebastiani, J. Spirkova, S. Berneschi, M. Brenci, R. Calzolai, A. Chiasera, M. Ferrari, M. Montagna, G. N. Conti, S. Pelli, and G. C. Righini, "Characterization of erbium doped lithium niobate crystals and waveguides," Opt. Mater. (Amst). **28**, 1292–1295 (2006).

12. H. C. Frankis, H. M. Mbonde, D. B. Bonneville, C. Zhang, R. Mateman, A. Leinse, and J. D. B. Bradley, "Erbium-doped TeO2 -coated Si3N4 waveguide amplifiers with 5 dB net gain," Photonics Res. **8**, 127 (2020).